\documentclass[12pt]{article}
\usepackage[utf8]{inputenc}
\usepackage[margin=1in]{geometry}
\usepackage{graphicx}
\usepackage{amsmath,amssymb}
\usepackage[colorlinks=true,linkcolor=blue,urlcolor=blue,citecolor=blue]{hyperref}
\usepackage{authblk}

\title{\vspace{-3mm} \bf Dark matter from mediator decay in early matter domination}
\author[1]{Rouzbeh Allahverdi \thanks{\href{mailto:rouzbeh@unm.edu}{rouzbeh@unm.edu}}}
\author[1]{Ngo Phuc Duc Loc \thanks{\href{mailto:locngo148@gmail.com}{locngo148@gmail.com}}} 
\author[2]{Jacek K.~Osi{\'n}ski \thanks{\href{mailto:jaksaosinski@gmail.com}{jaksaosinski@gmail.com}}}
\date{\vspace{-1.5cm}}
\affil[1]{\small Department of Physics and Astronomy, University of New Mexico, Albuquerque, NM 87131, USA}
\affil[2]{\small AstroCeNT, Nicolaus Copernicus Astronomical Center of the Polish Academy of Sciences, ul.~Rektorska 4, 00-614 Warsaw, Poland}

\begin{document}

\maketitle

\begin{abstract}
We study dark matter production from mediator decays in scenarios with an epoch of early matter domination. Particles that mediate interactions between dark matter and the standard model particles are kinematically accessible to the thermal bath as long as their mass is below the reheating temperature of the Universe after inflation. Decay of on-shell mediators can then lead to copious production of dark matter during early matter domination or a preceding radiation-dominated phase. In particular, for mediators that are charged under the standard model, it can exceed the standard freeze-in channel due to inverse annihilations at much lower temperatures (often by many orders of magnitude). The requirement to obtain the correct relic abundance severely constrains the parameter space for dark matter masses above a few TeV. 
%except for extremely heavy mediators with a mass larger than $10^{\rm 10}$ GeV. 
%dark matter except for mediator masses (well) above %$10^{10}$ GeV or prolonged epochs of EMD that possibly %possibly last until the onset of big bang nucleosynthesis.          

\end{abstract}

%\section{an outline for us to be organized}
%\begin{itemize}
%    \item describe the model 
%    \item describe possible EMD scenarios (moduli, other 
% heavy particles...) in addition to our scenario (X in 
% bath decaying to DM) 
%    \item results -- describe DM reaching/not reaching 
% equilibrium from decay, TO - mX figures, subdominance of 
% scattering (details can be in appendix) 
%    \item subsection on HS case - mX no longer has to be 
% above TeV 
%    \item 
%\end{itemize}

%%%%%%%%%%%%%%%%%%%%%%%%%%%%%%%%%%%%%%%%%%%%%%%%%%%%%%%%%%%
\section{Introduction}
While there are many lines of evidence for the existence of dark matter (DM) in the Universe~\cite{Bertone:2004pz}, the identity of DM is an outstanding problem at the interface of particle physics and cosmology. Explaining the observed DM abundance is another challenge that in addition depends on the details of the thermal history of the early Universe. 
Thermal freeze-out in a radiation-dominated (RD) Universe is a simple and attractive mechanism that can yield the correct relic abundance if the (thermally averaged) DM annihilation rate takes the specific value $\langle \sigma_{\rm ann} v \rangle = 3 \times 10^{-26}$cm$^3$ s$^{-1}$. However, in nonstandard cosmological histories, the correct DM abundance can be obtained for much larger or smaller values of $\langle \sigma_{\rm ann} v \rangle$~\cite{Kamionkowski:1990ni,Allahverdi:2020bys}. 

It is known that well-motivated classes of models arising from string theory generically lead to nonstandard histories that involve one or more epochs of early matter domination (EMD)~\cite{Kane:2015jia}. In general, an EMD phase arises when a matter-like component dominates the energy density of the Universe and eventually decays to establish a RD Universe prior to big bang nucleosynthesis (BBN). The matter equation of state can be due to coherent oscillations of a scalar field, like a string modulus, that is displaced from the minimum of its potential during inflation~\cite{Acharya:2008bk,Allahverdi:2013noa,Allahverdi:2014ppa,Cicoli:2016olq}.
% ,aparicio2015nonthermal,aparicio2016light,Cicoli_2022}
It can also arise from long-lived nonrelativistic quanta produced in the postinflationary Universe~\cite{Dror:2016rxc,Berlin:2016gtr,Dror:2017gjq,Cirelli:2018iax}. In fact, it has been recently shown that a visible-sector long-lived particle (LLP) with a weak-scale mass may achieve this~\cite{Allahverdi:2021grt,Allahverdi:2022zqr}. 

The indirect DM searches have set stringent constraints on the DM annihilation rate, most notably the recent Fermi-LAT results from dwarf spheroidal galaxies~\cite{Fermi-LAT:2015att} and Milky Way satellites~\cite{Fermi-LAT:2016uux}. An analysis of these results has pushed the upper limit on $\langle \sigma_{\rm ann} v \rangle$ below the nominal value of $3 \times 10^{-26}$ cm$^3$ s$^{-1}$ for DM masses below 20 GeV in a model-independent way~\cite{Leane:2018kjk} (larger masses can be excluded for specific annihilation channels). The increasingly tighter experimental bounds therefore warrant studying cases with a small DM annihilation rate. If $\langle \sigma_{\rm ann} v \rangle$ is very small, inverse annihilation will never be in thermal equilibrium and DM production occurs in the freeze-in regime. 

Freeze-in production during EMD has been studied in the 
literature~\cite{Hall:2009bx,Giudice:2000ex,Erickcek:2015jza,Calibbi:2021fld,Ahmed:2022tfm} (for a detailed review, see~\cite{Bernal:2017kxu}). These studies have mostly focused on the latter part of an EMD epoch where evolution of radiation is nonadiabatic. It has been noted that earlier stages of the thermal history can significantly contribute to freeze-in production and thereby dominate the DM relic abundance (for example, see~\cite{Allahverdi:2019jsc}). This is an important consideration because an epoch of EMD is typically preceded by other phases in the postinflationary history (notably the RD phase established at the end of inflationary reheating), especially for the last epoch of EMD that can end just before BBN.

A period of EMD starting with a non-negligible initial amount of radiation
%at its onset 
%that is preceded by a RD universe 
%typically starts with 
goes through an adiabatic phase during which $T \propto H^{2/3}$, as opposed to $T \propto H^{1/4}$ in the nonadiabatic phase. In fact, as we will see, the Universe is generally in this adiabatic phase for most of the EMD period, in terms of the temperature evolution. The faster redshift of temperature in this phase then implies that particles that mediate DM interactions with the standard model (SM) particles can be kinematically accessible to the thermal bath during EMD (or in a preceding phase) even if they are very heavy. The mediator particles can indeed reach equilibrium if their couplings to the SM particles are not too small, and hence their decay can be an important source of DM production. 
  
In this work, we present a detailed study of the contribution from on-shell mediator decays to the DM relic abundance in the freeze-in regime. By paying special attention to the adiabatic phase of EMD, we will show that these decays can easily dominate over the standard freeze-in production from inverse annihilations at much lower temperatures by many orders of magnitude. In fact, mediator decays alone can easily lead to DM overproduction in large parts of the parameter space. As we will see, the parameter space is tightly constrained for DM masses above a few TeV. 
%values of DM mass unless the mediators are (much) heavier %than $10^{10}$ GeV or for very long epochs of EMD that %could last until the onset of BBN. 
The resulting constraints are milder and the allowed parameter space will extend to smaller DM masses if DM and the mediator belong to a hidden sector with its own gauge symmetry.     

The rest of this paper is organized as follows. In Section~\ref{sec1}, we introduce the general picture for mediator coupling to DM and SM particles. In Section~\ref{sec3}, we discuss some important details of EMD epochs. In Section~\ref{sec4}, we present our main results and identify the allowed region of the parameter space in order for mediator decays not to overproduce DM. We also comment on how our results are modified in the case of hidden sector DM models. We conclude the paper in Section~\ref{sec5}. Some details of our calculations are discussed in the Appendix. 
  
%%%%%%%%%%%%%%%%%%%%%%%%%%%%%%%%%%%%%%%%%%%%%%%%%%%%%%%%%%%
\section{The set up}\label{sec1}

The main particles that are of interest in this work are the DM candidate $\chi$ and the mediator of its interactions with the SM particles $X$. Our focus is on DM production from $X$ decay, and hence we consider dimension-four operators that are linear in $X$ and involve $\chi$ and the SM particles (collectively denoted by $\psi$): 
%are possible: 
%
\begin{equation} \label{ints}
h_1 {\cal O}_4(X \chi \psi) ~ ~ ~ ~ ~ , ~ ~ ~ ~ ~ h_2 {\cal O}_4 (X \chi \chi) ~ ~ ~ ~ ~ , ~ ~ ~ ~ ~ h_3 {\cal O}_4 (X \psi \psi) .
\end{equation}
Here, $h_i$ are dimensionless couplings. We assume that $X$ is charged under the SM gauge group, while $\chi$ may or may not be a SM singlet. 
%remain agnostic about the bosonic or fermionic nature of it %and $\chi$, as well as the Lorentz structure of the %interaction terms. 
All operators in Eq.~(\ref{ints}) are invariant under the SM gauge symmetry and other symmetries that a given model may possess (for example, the symmetry that is responsible for stability of DM). For clarity, we consider some specific forms of operators in Eq.~(\ref{ints}) that arise in well-motivated particle physics models:
\vskip 2mm
\noindent
{\bf 1 -} $X$ a scalar and $\chi$ a fermion. In this case, one can have the following types of interactions (using the four-component notation for fermions):
\begin{equation} \label{ints1}
h_1 X {\bar \chi} (a_1 + b_1 \gamma^5) \psi + {\rm h.c.} ~ ~ , ~ ~ h_2 X {\bar \chi} (a_2 + b_2 \gamma^5) \chi + {\rm h.c.} ~ ~ , ~ ~ h_3 X {\bar \psi} (a_3 + b_3 \gamma^5) \psi + {\rm h.c.} ,
\end{equation}
where $\psi$ are the SM fermions. 

Supersymmetric extensions of the SM with $R$-parity conservation give rise to the first term in Eq.~(\ref{ints1}), while $R$-parity implies $a_2 = b_2 = a_3 = b_3 = 0$. In these models, the lightest supersymmetric particle (LSP) is stable and hence a DM candidate (for a review, see~\cite{Jungman:1995df}). Then $X$ can be any scalar superpartner that is coupled to the neutralino DM. We note that both $X$ and $\chi$ are charged under the SM in this case.

Another example is the model with GeV-scale DM proposed in~\cite{Allahverdi:2013mza}, which gives rise to the first and third terms in~(\ref{ints1}) with $a_1 = b_1 = a_3 = b_3 = 1/2$. In this case $X$ is a color-triplet and iso-singlet scalar that is coupled to the up-type and down-type RH quarks (as well as DM), and $\chi$ is a SM singlet fermion. %\footnote{In the supersymmetric version of this %model~\cite{BMN}, DM can be the scalar partner of $\chi$.}.

%Another possibility is that the DM candidate $\chi$ is a %scalar itself. In this case This can happen in the model %in~\cite{BMN} where the LSP is the scalar component of a %supersymmetric multiplet that is a singlet under the SM gauge %group. 
%
\vskip 1.5mm
\noindent
{\bf 2 -} $X$ a gauge boson and $\chi$ a fermion. In this case, one can have interactions of the following type:
\begin{equation} \label{ints2}
h_1 X_\mu {\bar \chi} (a_1 + b_1 \gamma^5) \gamma^\mu \psi + {\rm h.c.} ~ ~ , ~ ~ h_2 X_\mu {\bar \chi} (a_2 + b_2 \gamma^5) \gamma^\mu \chi + {\rm h.c.} ~ ~ , ~ ~ h_3 X_\mu {\bar \psi} (a_3 + b_3 \gamma^5) \gamma^\mu \psi + {\rm h.c.} ,
\end{equation}
with $\psi$ denoting the SM fermions. This happens, for example, when the SM gauge symmetry is extended and $X$ is 
%interactions of the $X \chi \chi$ and $X \psi \psi$ types can arise with $X$ being 
a gauge boson of the new symmetries. An example is the $U(1)_{B-L}$ extension of the SM~\cite{Mohapatra:1980qe} (also see~\cite{Davidson}) that leads to the second and third terms in~(\ref{ints2}) with $b_2 = b_3 = 0$. In this case, the $Z^{\prime}_{B-L}$ is coupled to all SM fermions as well as the RH neutrinos, the lightest of which can be the DM candidate~\cite{Burell:2011wh} \footnote{In the supersymetric $B-L$ model, the lightest RH sneutrino can be the DM particle (for example, see~\cite{Matchev,Allahverdi:2007wt,Arina}). In this case the relevant dimension-four operator has the form $i g_{B-L} X_\mu \chi^* \partial^\mu \chi + {\rm h.c.}$, where $g_{B-L}$ is the $U(1)_{B-L}$ gauge coupling.}. 
\vskip 2mm
%
%Interactions of the form $X \chi \chi$ and $X \psi \psi$ can %also arise if $X$ is a sufficiently heavy $R$-parity even %particle. 
%In models where the SM gauge symmetry is extended, %interactions of the $X \chi \chi$ and $X \psi \psi$ types can %arise with $X$ being a gauge boson of the new symmetries. For %example, in the $U(1)_{B-L}$ extension of the %SM~\cite{Mohapatra:1980qe} (also see~\cite{Davidson}), %$Z^{\prime}_{B-L}$ is coupled to all SM fermions as well as %the RH neutrinos the lightest of which can be a DM %candidate~\cite{Burell:2011wh}. 
%of a $U(1)$ extension of the SM with DM being the %lightest %higgsino in the $U(1)$ sector. 
%Finally, in minimal extensions of the SM that only include %new fields, interactions of the type $X \chi \psi$ and $X %\psi \psi$ can arise. For example, in the model with GeV-%scale DM proposed in~\cite{Allahverdi:2013mza}, $X$ is a %color-triplet and iso-singlet scalar that is coupled to the %up-type and down-type RH quarks and DM.
   
Eqs.~(\ref{ints1},\ref{ints2}) are meant as illustrative examples, rather than a full list, of dimension-four operators in Eq.~(\ref{ints}). The important conclusion is that, when $m_\chi \ll m_X$, the first two operators in Eq.~(\ref{ints}) generally result in the following partial decay width:
\begin{equation} \label{Xdec}
\Gamma_{X \rightarrow \chi} \simeq C_1 \times 
{h^2 \over 8 \pi} m_X ,
\end{equation}
where $h^2 = h^2_1 + h^2_2$, and $C_1$ is a factor whose exact value depends on the nature of $X$ and $\chi$ and the details of their couplings. In the following, we will take $C_1 \sim 1$. Given that $X$ is in equilibrium with the thermal bath at $T \geq m_X$, because of its SM charges, its decay can lead to copious production of DM even if $h$ is very small.  

The interactions in Eq.~(\ref{ints})
%, along with a possible $X \psi \psi$ coupling, 
also lead to annihilation/inverse annihilation $\chi \chi \leftrightarrow \psi {\bar \psi}$ and scattering $\chi \psi \leftrightarrow \chi \psi$ processes mediated by $X$. The former contributes to $\chi$ production from the thermal bath. However, $X$ decay typically dominates $\chi$ production as long as the Universe starts at a temperature $T \gg m_X$ so that on-shell $X$ particles exist in sufficient abundance. This is because $\Gamma_{\rm ann} \ll \Gamma_{X \rightarrow \chi}$, where $\Gamma_{\rm ann} = \langle \sigma_{\rm ann} v \rangle n_\chi$. 

We note that if $\Gamma_{\rm ann} \ll H(T \sim m_{\chi})$, the comoving number density of DM particles produced from $X$ decay will remain frozen. The annihilation rate at energies $E \sim m_\chi$ depends on the dimension of the effective operator $\chi \chi \psi \psi$ and its Lorentz structure. For example, assuming $S$-wave dominance, at $E \ll m_X$ we have: 
%dimension-5 and dimension-6 operators (arising, for %example, if $X$ is a fermion or a scalar/vector %respectively) 
%we have:
%
\begin{eqnarray} \label{xsection1}
\langle \sigma_{\rm ann} v \rangle_{{\rm dim}-5} & \simeq & C_2 \times {h^4 \over 16 \pi m^2_X} \, , \\
\langle \sigma_{\rm ann} v \rangle_{\rm dim-6} & \simeq & C_2 \times {h^4 E^2 \over 16 \pi m^4_X} \, , \label{xsection2}
\end{eqnarray}
where $C_2$ is a multiplicity factor. 

One final comment is in order. In general, nonrenormalizable operators mediated by $X$ can significantly contribute to production of DM in the freeze-in regime. Such contributions, however, are typically subdominant to $X$ decay due to the higher order of corresponding operators (see Appendix \ref{App:B}). Our main goal here is to demonstrate that mediator decays set tight constraints on the parameter space in order to not overproduce DM. Including additional contributions can further restrict the allowed regions of the parameter space.

%%%%%%%%%%%%%%%%%%%%%%%%%%%%%%%%%%%%%%%%%%%%%%%%%%%%%%%%%%%%
\section{Early matter domination: generalities}\label{sec3}

In this work, we consider EMD that starts at some point after the completion of inflationary reheating. The matter component $\phi$ carries an energy density $\rho_\phi$, and its decay at the rate $\Gamma_\phi$ feeds the radiation energy density $\rho_{\rm r}$. The evolution of $\rho_{\rm r}$ and $\rho_\phi$ is governed by the following system of Boltzmann equations:
\begin{eqnarray} \label{Boltzmann1}
{d \rho_{\rm r} \over dt} + 4 H \rho_{\rm r} & = & \Gamma_\phi \rho_\phi \, , \nonumber \\
{d \rho_{\rm \phi} \over dt} + 3 H \rho_{\rm \phi} & = & - \Gamma_\phi \rho_\phi \, , 
\end{eqnarray}
where
%
%\begin{equation}
the Hubble rate $H$ is set by the total energy density 
$\rho_{\rm tot} = \rho_\phi + \rho_{\rm r}$. 
%+ \rho_{\rm %rest}$. 
%\over 3 M^2_{\rm P}} .   
%\end{equation}
%
%Here, we remain agnostic about the exact content of the %universe before EMD and assume that it consists of a matter %component $\phi$ and radiation plus some other component %that makes up the rest. Note that as long as the latter has %an equation of state with $0 < w \leq 1$, then $\phi$ will %eventually dominate the energy density if it is %sufficiently long-lived. 

%$\rho_\phi (t) \propto a^{-3}(t)$ as opposed to $\rho_{\rm %r} \propto a^{-4}(t)$ during RD, with $a(t)$ being the %scale factor. This implies that $\phi$ eventually takes %over if it is sufficiently long-lived thereby giving rise %to a period of EMD. 

%\subsection{Stages of early matter domination}

We take the moment when $\rho_\phi = \rho_{\rm tot}/2$ to be the onset of the EMD epoch. We denote the Hubble rate and the temperature at this time by $H_{\rm O}$ and $T_{\rm O}$ respectively. Given that $\rho_{\rm r} = \rho_{\rm tot}/2$, we have:
\begin{equation} \label{HT}
H_{\rm O} = \left({\pi^2 g_{*{\rm O}} \over 45}\right)^{1/2} \left({T^4_{\rm O} \over M^2_{\rm P}}\right)^{1/2} ,  
\end{equation}
where \(M_{\rm P}\) is the reduced Planck mass and \(g_*\) counts the relativistic degrees of freedom at the time indicated by the subscript. 
Note that in the most extreme case we have $T_{\rm O} \simeq T_{\rm reh}$ (where $T_{\rm reh}$ is the inflationary reheat temperature). The EMD epoch that starts at $H \simeq H_{\rm O}$ consists of two phases:
\vskip 2mm
\noindent{\bf Adiabatic phase -} In this phase, the initial radiation dominates over that produced from $\phi$ decay. As a result, $T \propto a^{-1}$, with \(a\) being the scale factor, similar to RD even though the Universe is in an EMD epoch where $a \propto t^{2/3}$. During this stage, the relation between $H$ and $T$ follows:
\begin{equation} \label{A} 
H \simeq \left({\pi^2 g^{3/4}_* g^{1/4}_{*{\rm O}} \over 90}\right)^{1/2} {T^{3/2} {T^{1/2}_{\rm O}} \over M_{\rm P}}. 
\end{equation}
This phase eventually ends when the temperature drops to the following value:
\begin{equation} \label{Ttr}
T_{\rm tr} \simeq \left({g^4_{*{\rm R}} g_{*{\rm O}} \over g^5_*}\right)^{1/20} (T^4_{\rm R} T_{\rm O})^{1/5} . 
\end{equation}
\vskip 1.5mm
\noindent 
{\bf Nonadiabatic phase -} There is a transition to the nonadiabatic phase at $T_{\rm tr}$ where radiation produced by $\phi$ decay becomes dominant over the preexisting amount. During this stage, we have:
\begin{equation} \label{NA}
H \simeq {5 g_* \over 6 g^{1/2}_{*{\rm R}}} \left({\pi^2 \over 30}\right)^{1/2} \left({T^4 \over T^2_{\rm R} M_{\rm P}}\right) ,
\end{equation}
which implies that $T \propto a^{-3/8}$. This stage extends all the way from $T \simeq T_{\rm tr}$ to the end of the EMD epoch when $H \sim \Gamma_\phi$ and $T \simeq T_{\rm R}$. 
\vskip 2mm
This discussion holds even if some non-radiation component with an equation-of-state parameter $0 < w \leq 1$ makes the dominant contribution to $\rho_{\rm tot}/2$ at the onset of EMD, provided that it is not converted into radiation. Note that a component with $w > 0$ is redshifted faster than matter, and hence will become increasingly subdominant during EMD without any need to decay to radiation. If it feeds radiation, on the other hand, entropy will increase at a faster rate during EMD thereby leading to a shorter adiabatic phase and a higher $T_{\rm tr}$ than what is mentioned above.    

%\subsection{Examples of early matter domination scenarios}

To illustrate the relative duration of the two stages of EMD, we consider two generic possibilities regarding the origin of such an epoch: coherent oscillations of a scalar field that is displaced from the minimum of its potential, and nonrelativistic particles produced from the thermal bath.
\vskip 2mm
\noindent
{\bf (1) Coherent oscillations -} Consider a scalar field $\phi$ with mass $m_\phi$ (a notable example is string moduli~\cite{Kane:2015jia}). If $m_\phi \ll H_{\rm inf}$, where $H_{\rm inf}$ is the Hubble rate during inflation, then $\phi$ can be significantly displaced from the minimum of its low-energy potential during inflation. It typically starts to oscillate about this minimum when $H \simeq m_\phi$ and its coherent oscillations behave like non-relativistic particles of mass $m_\phi$. If the Universe is RD at this time, and the initial fractional energy density of $\phi$ is $\alpha$, coherent oscillations start to dominate when the Hubble expansion rate is:
\begin{equation} \label{dom1}
H_{\rm O} \simeq \alpha^2 m_\phi .
\end{equation}
This signals the start of an EMD epoch when the temperature reaches the following value:
\begin{equation} \label{Onset}
T_{\rm O} \simeq \alpha {\left(45 \over \pi^2 g_{*{\rm O}}\right)}^{1/4} {(m_\phi M_{\rm P})}^{1/2} .  
\end{equation}
On the other hand, if the Universe is still dominated by inflaton oscillations when $H \sim m_\phi$, we will instead have:
\begin{equation} \label{dom2}
H_{\rm O} \simeq \alpha^2 H_{\rm reh} ,
\end{equation}
where $H_{\rm reh} < m_\phi$ is the Hubble rate when a RD Universe is established after inflationary reheating. Therefore, given that $\alpha \lesssim 1$ \footnote{In realistic situations, we typically have $\alpha \gtrsim {\cal O}(10^{-2})$~\cite{Cicoli:2016olq}.}, the temperature at the onset of the EMD epoch follows:
\begin{equation} \label{TO1}
T_{\rm O} \lesssim {\left(45 \over \pi^2 g_{*{\rm O}}\right)}^{1/4} {(m_\phi M_{\rm P})}^{1/2} .     
\end{equation}
\vskip 1.5mm
\noindent
{\bf (2) Nonrelativistic quanta -} Consider a (bosonic or fermionic) particle $\phi$ from decay or scattering processes in the thermal bath. The produced quanta become nonrelativistic at $T \sim m_\phi$. If the fractional energy density of $\phi$ at this time is $\alpha$, its quanta dominate the Universe when the Hubble expansion rate is:
\begin{equation}
H_{\rm O} \sim \alpha^2 \left({\pi^2 g_{*m_\phi} \over 45}\right)^{1/2} {m^2_\phi \over M_{\rm P}} .    
\end{equation}
We note that $\alpha \lesssim g^{-1}_{*m_\phi}$, with the maximum occurring when $\phi$ reaches thermal equilibrium. Thus, in this case, we have:
\begin{equation} \label{TO2}
T_{\rm O} \lesssim \left({g^3_{*m_\phi} \over g_{*{\rm O}}}\right)^{1/2} m_\phi .    
\end{equation}
\vskip 2mm
To get a feeling of the relative duration of the adiabatic and nonadiabatic phases, let us consider the $m_\phi \simeq 10^2-10^6$ GeV mass range. The upper and lower limits correspond to the cases when $\phi$ is a string modulus~\cite{Allahverdi:2013noa,Allahverdi:2014ppa} and a weak-scale visible sector particle~\cite{Allahverdi:2021grt,Allahverdi:2022zqr} respectively. Eqs.~(\ref{TO1},\ref{TO2}) then give us:
\begin{eqnarray} \label{TOmax}
&& T_{\rm O} \sim 10^{10}-10^{12} ~ {\rm GeV} ~ ~ ~ ~ ~ ({\rm Case} ~ 1) \, , \nonumber \\
&& T_{\rm O} \sim 1-10^{4} ~ {\rm GeV} ~ ~ ~ ~ ~ ~ ~ ~ ~ ~ ({\rm Case} ~ 2) \, ,
\end{eqnarray}
where we have used $g_* \sim {\cal O}(100)$ as in the case of the SM. We therefore see that $T_{\rm O}$ can be within a vast range spanning many orders of magnitude. 

The important point is that $T_{\rm tr}$ is in general not extremely larger than $T_{\rm R}$. It follows from Eq.~(\ref{Ttr}) that $T_{\rm tr}/T_{\rm R}$ scales as $(T_{\rm O}/T_{\rm R})^{1/5}$. For example, for $T_{\rm R} \sim 10$ MeV, we find:
\begin{eqnarray} \label{Ttrmax}
&& T_{\rm tr} \sim 100-300 ~ {\rm GeV} ~ ~ ~ ~ ~ ({\rm Case} ~ 1) \, , \nonumber \\
&& T_{\rm tr} \sim 0.03-10 ~ {\rm GeV} ~ ~ ~ ~ ~ ~ ({\rm Case} ~ 2) \, ,    
\end{eqnarray}
We see that while $T_{\rm O}$ is typically much higher in case 1, see Eq.~(\ref{TOmax}), in both cases the majority of the EMD epoch is spent in the adiabatic phase $T_{\rm tr} \lesssim T \lesssim T_{\rm O}$.

%%%%%%%%%%%%%%%%%%%%%%%%%%%%%%%%%%%%%
\section{Results}\label{sec4}

We consider the evolution of the \(\chi\) number density produced from the decay of \(X\) in the bath, as described by the following Boltzmann equation~\cite{Hall:2009bx,Calibbi:2021fld,DEramo:2018vss,Du:2021jcj}:
\begin{equation}\label{eq:Boltzmann_chi}
\frac{dn_\chi}{dt} + 3Hn_\chi = \Gamma_{X\to\chi} \frac{n_{X}^{\rm eq}}{n_{\chi}^{\rm eq}} \frac{K_1(m_X/T)}{K_2(m_X/T)} (n_{\chi}^{\rm eq} - n_\chi) \,,
\end{equation}
where \(K_i\) are the modified Bessel functions of the second kind. The superscript `eq' denotes the corresponding equilibrium number density. Note that $X$ reaches equilibrium with the thermal bath at $T \gg m_X$ due to its SM charges. $n_X$ will follow its equilibrium value at $T < m_X$ if annihilation to the SM particles is efficient or in the presence of an efficient decay mode via the $h X \psi \psi$ term in Eq.~(\ref{ints}). 

In general, the \(\chi\) number density will become frozen at some time in the cosmological history. Depending on the values of the parameters, this can occur in any of the phases described in the previous section. Furthermore, the freezing process may proceed as a freeze-in, where \(\chi\) never achieves equilibrium with the bath, or a freeze-out, where the production rate is sufficient to reach equilibrium, with \(\chi\) subsequently decoupling. 

We can analytically estimate the abundance of \(\chi\) produced from \(X\) decay, expressed as the number density normalized by the entropy density \(n_\chi / s\), in the following way (see Appendix \ref{App:A} for details). We first define the temperature of the background bath at the time when the \(\chi\) number density freezes as \(T_{\rm f}\). Beyond this point the number density is only redshifted with expansion as \(n_\chi \propto a^{-3}\). We express the number density \(n_\chi\) at this time as: 
\begin{equation}
n_\chi(T_{\rm f}) = F(\gamma_\chi) n_\chi^{\rm eq}(T_{\rm f}) ,
\end{equation} 
where the function \(F(\gamma_\chi)\) accounts for the possibility of not reaching equilibrium, with \(\gamma_\chi \equiv \Gamma_{X \to \chi} / H(T = m_X)\). 
% If equilibrium is reached, we have \(\gamma_\chi > 1\) and \(F(\gamma_\chi) = 1\), while if equilibrium is not reached, or \(\gamma_\chi \ll 1\), then the number density of \(\chi\) saturates to \(n_\chi \approx 7 \gamma_\chi n_\chi^{\rm eq}\) before \(T \approx m_X / 10\). 
We will typically consider \(T_{\rm f} \gg m_\chi\) such that the equilibrium number density above is given by the relativistic expression proportional to \(T_{\rm f}^3\). 

In general, the \(\chi\) number density produced from \(X\) decay can become frozen at different stages of the cosmological history described in the previous section: 
%described in the previous section, depending on the values %of the parameters. Additionally, this can occur as a %freeze-in process, where equilibrium is never achieved, or %a freeze-out where the production rate is sufficient to %reach equilibrium and \(\chi\) subsequently decouples.
\vskip 2mm
\noindent
{\bf (1)} $T_{\rm R} < m_X \lesssim T_{\rm tr}$ - In this case, $n_\chi$ freezes during the nonadiabatic phase of EMD, and the final DM abundance is given by:
\begin{equation} \label{nonadiabatic}
\left(\frac{n_\chi}{s}\right)^{\rm NA} \simeq 0.53 \frac{g_\chi g_{*\rm R}}{g_{*\rm f}^2} F(\gamma_\chi) \left(\frac{T_{\rm R}}{T_{\rm f}}\right)^5 ,
\end{equation}
% \begin{equation} \label{nonadiabatic}
% \left(\frac{n_\chi}{s}\right)_{\rm R}^{\rm nonad} = \frac{A^2 72\zeta(3)g_\chi g_{*\rm R}}{5\pi^4 g_{*\rm f}^2} F(\gamma_\chi) \left(\frac{T_{\rm R}}{T_{\rm f}}\right)^5 .
% \end{equation}
%
where $g_\chi$ is the number of degrees of freedom in $\chi$.  
\vskip 1.5mm
\noindent
{\bf (2) $T_{\rm tr} < m_X \lesssim T_{\rm O}$ -} In this case, $n_\chi$ freezes during the adiabatic phase of EMD and the DM relic abundance follows:   
\begin{equation} \label{adiabatic}
\left(\frac{n_\chi}{s}\right)^{\rm A} \simeq 0.28 \frac{g_\chi}{g_{*\rm f}} F(\gamma_\chi) \left(\frac{T_{\rm R}}{T_{\rm O}}\right) .
\end{equation}
% \begin{equation} \label{adiabatic}
% \left(\frac{n_\chi}{s}\right)_{\rm R}^{\rm ad} = \frac{45\zeta(3) g_\chi}{2\pi^4 g_{*\rm f}} F(\gamma_\chi) \left(\frac{T_{\rm R}}{T_{\rm O}}\right) .
% \end{equation}
%
\vskip 1.5mm
\noindent
{\bf (3) $T_{\rm O} < m_X \lesssim T_{\rm reh}$ -} In this case, $n_\chi$ freezes during the RD phase preceding the EMD epoch. Given that the frozen $n_\chi$ and $s$ are both redshifted $\propto a^{-3}$ during this RD phase and the adiabatic phase of EMD alike, we have:
\begin{equation} \label{radiation}
\left(\frac{n_\chi}{s}\right)^{\rm RD} \simeq 0.28 \frac{g_\chi}{g_{*\rm f}} F(\gamma_\chi) \left(\frac{T_{\rm R}}{T_{\rm O}}\right) .
\end{equation}
% \begin{equation} \label{radiation}
% \left(\frac{n_\chi}{s}\right)_{\rm R}^{\rm RD} = \frac{45\zeta(3) g_\chi}{2\pi^4 g_{*\rm f}} F(\gamma_\chi) \left(\frac{T_{\rm R}}{T_{\rm O}}\right) .
% \end{equation}
%
\vskip 2mm
If $\gamma_\chi \ll 1$, production of \(\chi\) will not be strong enough to reach equilibrium and the number density of \(\chi\) saturates to \(n_\chi \approx 7 \gamma_\chi n_\chi^{\rm eq}\). The temperature at the time of $\chi$ freeze-in in this case falls roughly within the range \(m_X / 10 \lesssim T_{\rm f} \lesssim m_X / 5\). On the other hand, if equilibrium is reached, we have \(\gamma_\chi \gtrsim 1\) and \(F(\gamma_\chi) = 1\), with the temperature at $\chi$ freeze-out not being limited to the previous range.

% If equilibrium is reached, we have \(\gamma_\chi > 1\) %and \(F(\gamma_\chi) = 1\), while if equilibrium is not %reached, or \(\gamma_\chi \ll 1\), then the number density %of \(\chi\) saturates to \(n_\chi \approx 7 \gamma_\chi %n_\chi^{\rm eq}\) before \(T \approx m_X / 10\). 

In the case that \(\chi\) reaches equilibrium in the nonadiabatic phase (or maintains it through the transition from the adiabatic phase), we can estimate the decoupling temperature using the RHS of the Boltzmann equation Eq.~\eqref{eq:Boltzmann_chi}: 
\begin{equation}\label{eq:decouple_NA}
\Gamma_{X\to\chi} \frac{g_X m_X^2}{g_\chi m_\chi^2} \frac{K_1(m_X/T_{\rm f})}{K_2(m_\chi/T_{\rm f})} 
\approx 3 H(T_{\rm f}) \,,
\end{equation}
where we have made use of \(n^{\rm eq}_{\chi}(T) = (g_{\chi} / 2\pi^2) m_{\chi}^2 T K_2(m_{\chi}/T)\), and similarly for \(n^{\rm eq}_X(T)\)~\footnote{Given that $g_\chi = 1$ (3/4) for a single bosonic (fermionic) degree of freedom, we take $g_\chi = 1$ in our numerical calculations (and similarly for $g_X$).}. 
Using this to find \(T_{\rm f}\), which we identify as the decoupling temperature in this case,
% we can replace \(T_{\rm f}\) with \(T_{\rm d}\) in %Eq.~\eqref{eq:n/s_NA}. This will give 
we can obtain the abundance at the end of EMD \((n_\chi/s)_{\rm R}\) for cases where \(\chi\) remains relativistic and in equilibrium in the nonadiabatic phase at temperatures below \(m_X\). We note that if equilibrium is reached during the adiabatic phase, but equilibrium is not maintained through the transition (i.e., decoupling occurs before \(T_{\rm tr}\)), then there is no dependence on the decoupling temperature, as seen in Eq.~\eqref{adiabatic} above. 

% -- mention Tf relation to mX if freeze-in, and decoupling for freeze-out 

% -- no dependence on decoupling temp if it is larger than Ttr

% -- mention other decoupling cases? 

With \(T_{\rm f}\) determined, 
%from Eq.~(\ref{eq:decouple_NA})
%for both freeze-in and freeze-out scenarios, 
we can now compare the frozen \(\chi\) abundance to the observed DM relic abundance in order to constrain such scenarios: 
\begin{equation}
\frac{n_\chi}{s} \lesssim \left(\frac{n_\chi}{s}\right)_{\rm obs} \simeq 4 \times 10^{-10} \times  {1 ~ {\rm GeV} \over m_\chi} \, .
\end{equation}
Here, we only require that DM is not overproduced from $X$ decays. Underproduction may be compensated for by other sources that create $\chi$ (for example, direct decay of $\phi$ or inverse annihilations).  

We numerically solve the Boltzmann equations for the background energy densities, given by Eq.~\eqref{Boltzmann1} 
% \begin{equation}
%     \frac{d\rho_{\rm r}}{dt} + 4H\rho_{\rm r} = \Gamma_\phi \rho_\phi
% \end{equation}
% 
% \begin{equation}
%     \frac{d\rho_\phi}{dt} + 3H\rho_\phi = - \Gamma_\phi \rho_\phi
% \end{equation}
together with the equation for the DM number density produced from \(X\) decay, Eq.~\eqref{eq:Boltzmann_chi}. We would like to emphasize that while $\Gamma_{X \rightarrow \chi}$ is essentially an input in~(\ref{eq:Boltzmann_chi}), it can be calculated within the particle physics models mentioned in Section \ref{sec1} according to Eq.~(\ref{Xdec}). In particular, the model in~\cite{Allahverdi:2022zqr} provides an explicit example where decay of a scalar $X$ is the main source of GeV-scale fermionic DM and also produces long-lived particles that are responsible for a period of EMD~\cite{Allahverdi:2021grt}. %\cite{Allahverdi:2013mza}
%Furthermore, given that $g_\chi = 1$ (3/4) for a single %bosonic (fermionic) degree of freedom, we take $g_\chi = g_X %= 1$ in our numerical calculations. 

% from \cite{DEramo:2018vss} Eqs.~3.1 and 3.4
% \begin{equation}\label{eq:Boltz}
%     \frac{dn_\chi}{dt} + 3Hn_\chi = \Gamma_{X\to\chi} \frac{n_{X,{\rm eq}}}{n_{\chi,{\rm eq}}} \frac{K_1(m_X/T)}{K_2(m_X/T)} (n_{\chi,{\rm eq}} - n_\chi)
% \end{equation}
% where \(K_i\) are the modified Bessel functions of the second kind. 
% 
% For code: 
% 
% \(T_{\rm O}\) gives initial \(\rho_\phi\), \(T_{\rm R}\) gives \(\Gamma_\phi\), and \(\left<\sigma v\right>\) (or \(h\)) determines strength of \(\Gamma_{X\to\chi}\) for a given \(m_X\). Initial \(\rho_{\rm r}\) given by arbitrary initial time. Can also include a scattering term. 
% 

\begin{figure}[t]
    \centering
    \includegraphics[width = 0.90\textwidth]{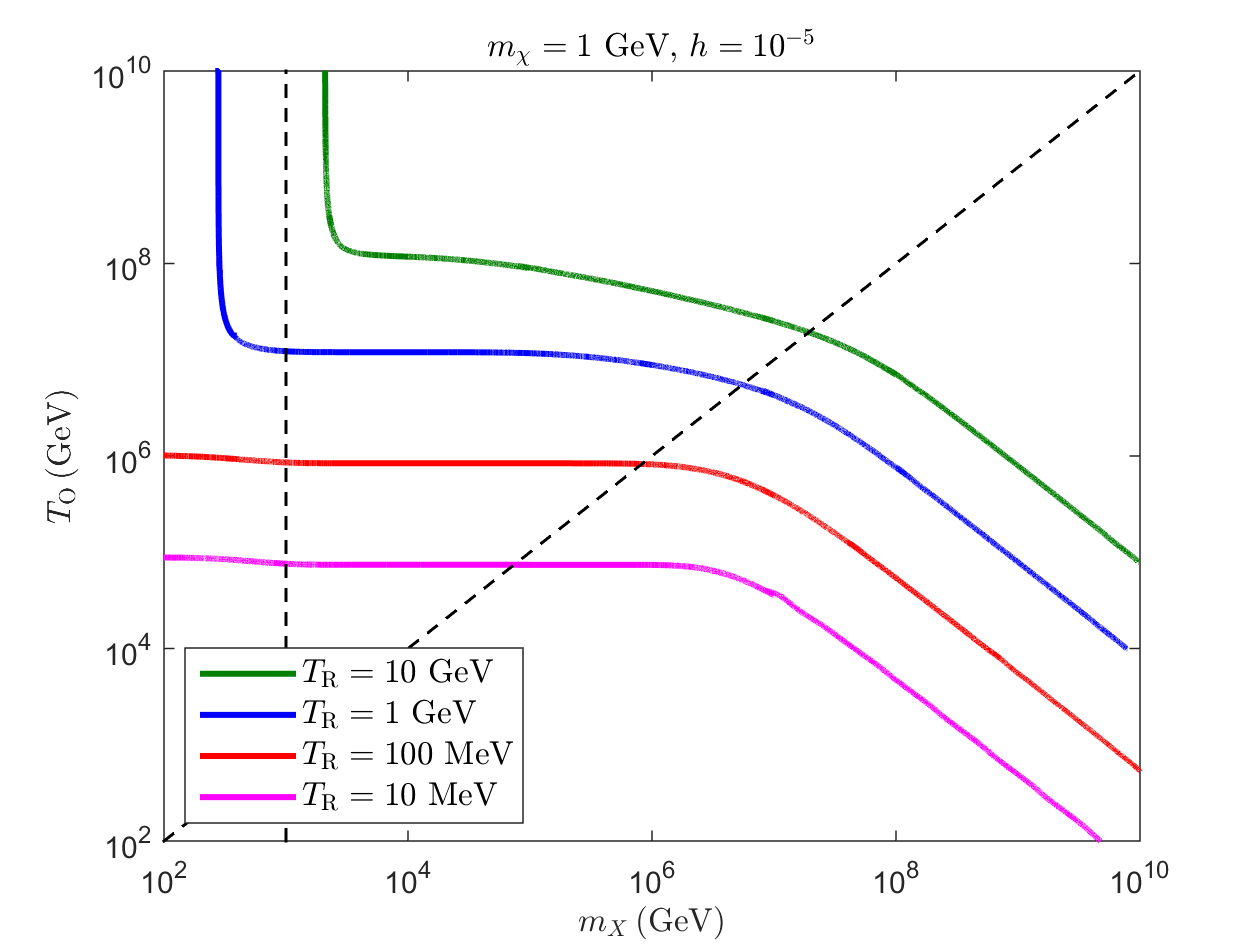}

    \caption{The \(T_{\rm O}\)-\(m_X\) plane for varied EMD end temperature \(T_{\rm R}\) with other parameters held fixed. Solid contours show parameter combinations that achieve the observed DM relic abundance. The region above (below) a given contour corresponds to underproduction (overproduction) and is therefore allowed (constrained). The diagonal dashed line shows \(T_{\rm O} = m_X\) for reference, while the vertical dashed line sits at \(m_X = 1\) TeV. 
    %For the cases that reach equilibrium, the transition %between horizontal/vertical occurs when the decoupling %temperature (not \(m_X\)) is roughly equal to \(T_{\rm %tr}\). 
    }
    \label{fig:TO_mX1}
\end{figure}

\begin{figure}[ht!]
    \centering
    \includegraphics[width = 0.90\textwidth]{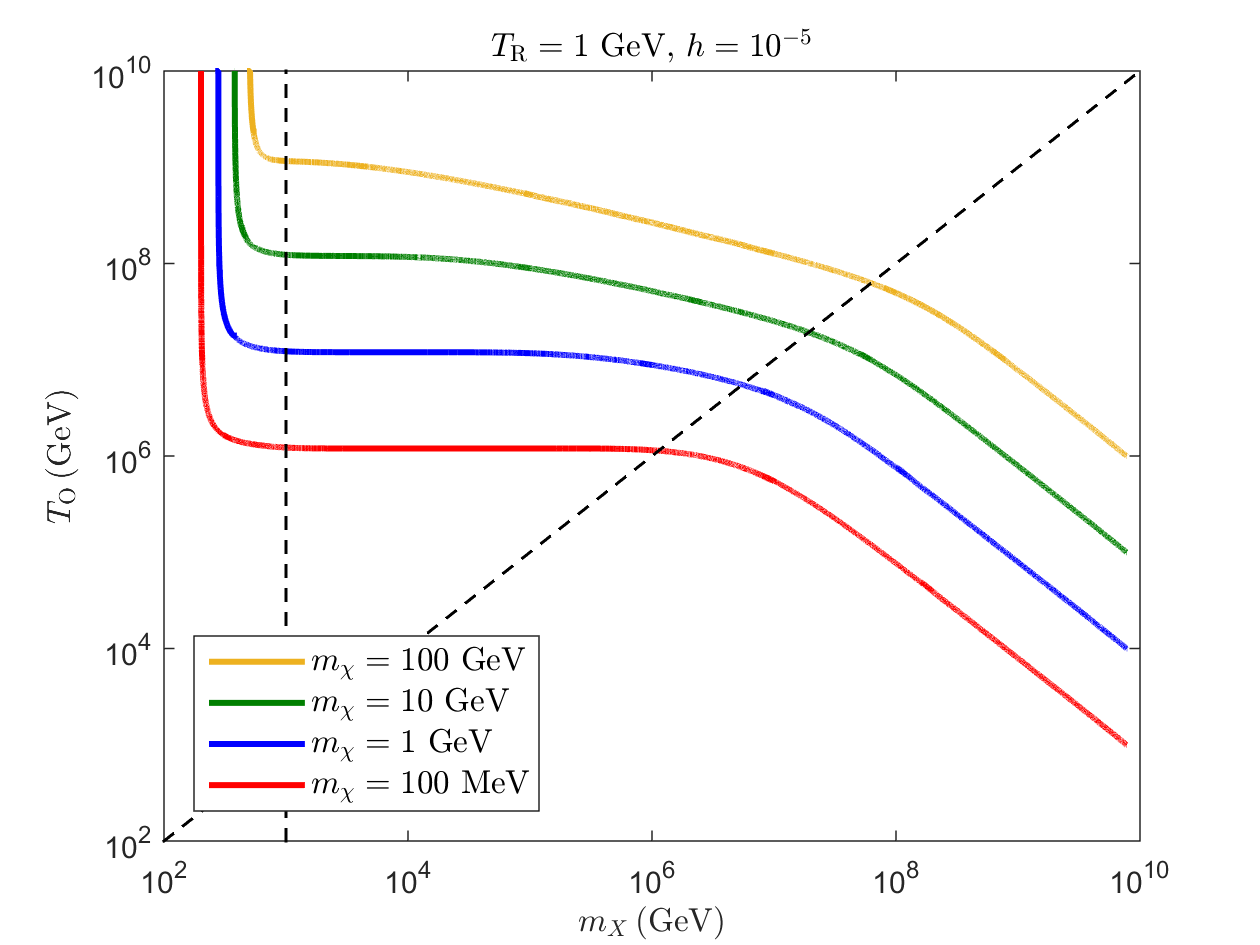}
    \caption{The \(T_{\rm O}\)-\(m_X\) plane for varied DM mass \(m_\chi\) with other parameters held fixed. As in Fig.~\ref{fig:TO_mX1}, solid contours correspond to the observed DM relic abundance, with the region above (below) a given contour being allowed (constrained). 
    %For the cases that reach equilibrium, the transition %between horizontal/vertical occurs when the decoupling %temperature (not \(m_X\)) is roughly equal to \(T_{\rm %tr}\). 
    }
    \label{fig:TO_mX2}
\end{figure}

\begin{figure}[ht!]
    \centering
    \includegraphics[width = 0.90\textwidth]{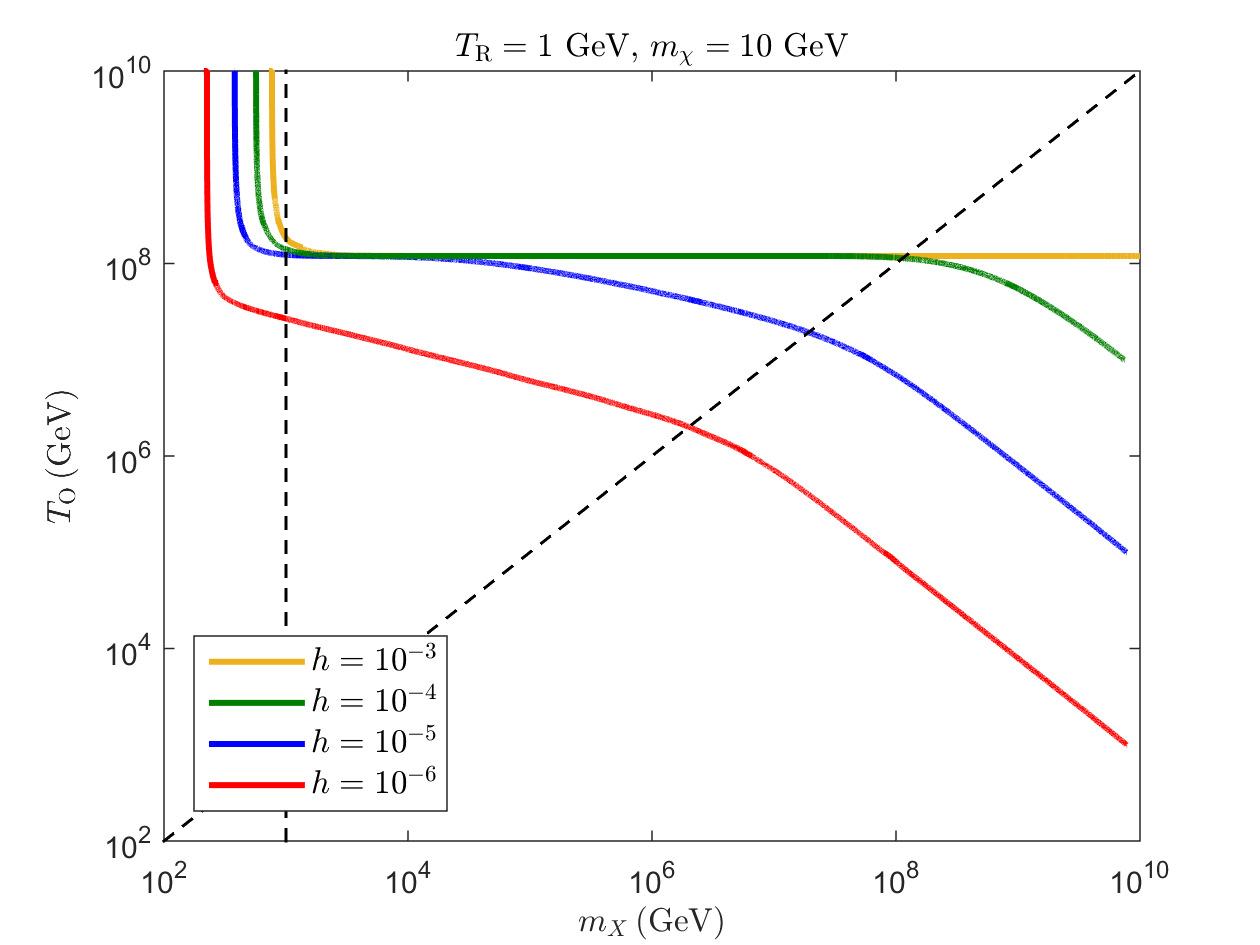}
    \caption{The \(T_{\rm O}\)-\(m_X\) plane for varied coupling \(h\) with other parameters held fixed. As in Fig.~\ref{fig:TO_mX1}, solid contours correspond to the observed DM relic abundance, with the region above (below) a given contour being allowed (constrained). 
    %For the cases that reach equilibrium, the transition %between horizontal/vertical occurs when the decoupling %temperature (not \(m_X\)) is roughly equal to \(T_{\rm %tr}\). 
    }
    \label{fig:TO_mX3}
\end{figure}

In Figs.~\ref{fig:TO_mX1}, \ref{fig:TO_mX2}, and \ref{fig:TO_mX3} we show contours in the \(T_{\rm O}\)-\(m_X\) plane that achieve the observed DM relic abundance for the values of the parameters shown in each figure. The region above and to the right of a given contour corresponds to underproduction of DM. In Fig.~\ref{fig:TO_mX1} we vary $T_{\rm R}$ for fixed $h$ and $m_\chi$. 
%two benchmarks with $m_\chi = 100$ GeV (right panel) and %$m_\chi = 1$ GeV (left panel) as representatives of weak %scale DM and light DM respectively. 
% vary \(T_{\rm R}\) for two benchmark points, 
In Fig.~\ref{fig:TO_mX2} we vary the DM mass \(m_\chi\) while keeping $h$ and $T_{\rm R}$ fixed. 
%as in Fig.~1 and two benchmarks with $T_{\rm R} = 1$ GeV %(right panel) and $T_{\rm R} = 10$ MeV representing %situations with EMD driven by a string modulus with %$m_\phi = {\rm few} \times 1000$ TeV and a very late EMD %epoch that lasts until the onset of BBN respectively. 
In Fig.~\ref{fig:TO_mX3} we vary $h$ for fixed values of $T_{\rm R}$ and $m_\chi$. 
%mentioned above in different panels. 

The vertical dashed line denotes the current LHC limit on new particles charged under the SM, which we loosely take to be $m_X \gtrsim {\cal O}({\rm TeV})$ \footnote{The exact bound depends on the SM charge assignment of $X$ (for example, colored versus colorless particles).}. One could also constrain $T_{\rm O}$ by considering the absolute upper bound $T_{\rm O} < T_{\rm reh}$ where $T_{\rm reh} \lesssim 10^{-1} (H_{\rm inf} M_{\rm P})^{1/2}$ \footnote{This upper bound is saturated in the case of instant reheating and thermalization. The exact value of $T_{\rm reh}$ depends on the details of reheating (for reviews, see~\cite{Allahverdi:2010xz,Amin:2014eta}) and can be much lower than this upper bound.}. The current Planck bounds~\cite{Planck:2018jri} on the tensor-to-scalar ratio correspond to $H_{\rm inf} \lesssim 10^{13}$ GeV, which results in a very weak limit $T_{\rm reh} \lesssim 10^{15}$ GeV. Other considerations can significantly tighten this bound. For example, the recently proposed trans-Planckian censorship conjecture of the swampland program sets an upper limit of $(H_{\rm inf} M_{\rm P})^{1/2} \lesssim 10^9$ GeV in order for modes with sub-Planckian wavelength to not exit the horizon during inflation. This would result in $T_{\rm O} < 10^9$ GeV thereby removing the very top of the $T_{\rm O}-m_X$ plane. Model-dependent bounds on $H_{\rm inf}$ lead to similar restrictions. 

The diagonal dashed line shows where $T_{\rm O} = m_X$. 
% Also note that we change the style from solid to dashed in the region below the $T_{\rm R} = m_X$ line. 
Extending the contours to the region below this line assumes that the Universe is in a RD phase from $T = T_{\rm O}$ all the way up to $T \simeq m_X$. This will not be the case if $m_X$ exceeds the reheating temperature after inflation $T_{\rm reh}$, or if there is another epoch of EMD at $H > H_{\rm O}$. Thus, the dependence on the details of the thermal history above $T_{\rm O}$ requires the $m_X > T_{\rm O}$ half-plane to be treated with some care. 
% Thus, to indicate of the dependence on the details of the thermal history above $T_{\rm O}$, we treat the $m_X > T_{\rm O}$ half-plane with some care.      

A general feature seen in all figures is that each contour starts as a vertical line at sufficiently small $m_X$.
%, and then turns to become (almost) horizontal. 
Along the vertical segments the bulk of DM production takes place in the nonadiabatic phase corresponding to $m_X \ll T_{\rm tr}$ where, see Eq.~(\ref{nonadiabatic}), the DM relic abundance is independent from $T_{\rm O}$. The vertical segment starts to turn at $m_X \sim T_{\rm tr}$. We note that in Fig.~\ref{fig:TO_mX2} the turning points are stacked on top of each other around values of $m_X$ within the same order of magnitude (unlike Fig.~\ref{fig:TO_mX1}). This is because $T_{\rm tr}$ is much more sensitive to $T_{\rm R}$ than $T_{\rm O}$ as can be seen in Eq.~(\ref{Ttr}). Since $T_{\rm R}$ is fixed in Fig.~\ref{fig:TO_mX2}, the change in $T_{\rm tr}$ for different contours is rather minimal, while this is not the case in Fig.~\ref{fig:TO_mX1}. The more significant dependence of $T_{\rm tr}$ on $T_{\rm R}$ also explains the apparently missing vertical segment in the contours for $T_{\rm R} = 10$ MeV and $100$ MeV in Fig.~\ref{fig:TO_mX1}. In this case, $T_{\rm tr}$ is smaller than the range of $m_X$ shown in that figure.

Another visible feature in all figures is the presence of a plateau beyond $m_X \sim T_{\rm tr}$ where the contours continue as a horizontal line for some range of $m_X$. Along this segment, DM production from $X$ decay occurs in the adiabatic phase. According to Eq.~(\ref{adiabatic}), the relic abundance has no dependence on $m_X$ as long as $F(\gamma_\chi) = 1$ (happening for $\gamma_\chi \gtrsim 1$). Note that in the adiabatic phase, see Eq.~(\ref{A}), $\gamma_\chi \propto (m_X T_{\rm O})^{-1/2}$. This implies that for sufficiently small values of $T_{\rm O}$, we have $\gamma_\chi \gtrsim 1$ all the way up to $m_X = T_{\rm O}$. Assuming a RD phase prior to EMD, $\gamma_\chi \propto m_X^{-1}$ for $m_X > T_{\rm O}$. It will therefore drop below one at some value of $m_X$, beyond which DM production is not in equilibrium and the contour has a negative slope. This behavior is seen for contours with smaller values of $T_{\rm R}$ ($m_\chi$) in Fig.~\ref{fig:TO_mX1} (Fig.~\ref{fig:TO_mX2}). Increasing $T_{\rm O}$ results in a smaller value for $\gamma_\chi$ during the EMD epoch, in which case DM production can drop out of equilibrium during the adiabatic phase of EMD.
%at $m_X < T_{\rm O}$. 
The contour then starts to fall 
%with two different slopes 
at $m_X < T_{\rm O}$ and keeps doing so for $m_X > T_{\rm O}$ (but with a different slope due to the different dependence of $\gamma_\chi$ on $m_X$). This is the behavior seen for contours with larger values of $T_{\rm R}$ ($m_\chi$) in Fig.~\ref{fig:TO_mX1} (Fig.~\ref{fig:TO_mX2}). 

% \textbf{clarify this last statement - the larger values of \(m_\chi\) and \(T_{\rm R}\) drop out of equilibrium in the adiabatic region, while the smaller ones drop in the assumed prior RD period. {\underline{R.A. I have modified the statement now. Please have a look.}}}

An additional feature seen in Fig.~\ref{fig:TO_mX3} is the asymptotic behavior of the contours as $h$ increases. In fact, the flat horizontal segment of contours with sufficiently large $h$ starts at the same $T_{\rm O}$ and $m_X$. 
%Further increasing $h$ only extends the plateau toward %larger $m_X$. 
This is because once $\gamma_\chi \sim {\cal O}(1)$ at $T_{\rm tr}$, $\chi$ reaches thermal equilibrium during the adiabatic phase of EMD. Further increasing $h$ only extends the plateau toward larger values of $m_X$ when $\gamma_\chi$ falls below one again.  
%represent different values of the coupling $h$ for fixed %$m_\chi$ and $T_{\rm R}$.

Given that the region below and to the left of each contour is ruled out as it gives rise to DM overproduction, we can draw some important conclusions about the allowed parts of the $T_{\rm O}-m_X$ plane:  
\vskip 2mm
\noindent
$\bullet$ The constraint $m_X \gtrsim {\cal O}({\rm TeV})$ does not severely restrict regions with DM underproduction unless $T_{\rm R} \ll 1$ GeV and/or $h \ll 10^{-5}$. It mainly affects regions where DM is overproduced, which are already ruled out. 
\vskip 1.5mm
\noindent
$\bullet$ The parameter space significantly opens up in the lower half-plane $T_{\rm O} < m_X$. However, in this part the main contribution to the DM relic abundance comes from the pre-EMD phase that requires detailed knowledge of the postinflationary thermal history.   
\vskip 1.5mm
\noindent
$\bullet$ Cases with $m_\chi > {\cal O}({\rm TeV})$ are essentially ruled out unless $T_{\rm O}$ and $m_X$ are (well) above $10^{10}$ GeV (see Fig.~\ref{fig:TO_mX2}). This is only marginally possible in models with large $T_{\rm O}$ (like modulus-driven EMD), and not viable in models with intermediate values of $T_{\rm O}$, see Eq.~(\ref{TOmax}).  % 
\vskip 2mm

%In general the contours have a few distinct behaviors that %can be understood as follows. Beginning at the bottom %right, the contours rise with a slope of ....... %corresponding to freeze-in during the RD period prior to %EMD, with \(m_X > T_{\rm O}\). This continues until roughly %\(m_X \approx T_{\rm O}\) at which point they begin to %flatten. 

%-- once TO goes below TR, no EMD, so expect contours to %become vertical again 

%-- include TR = 10 MeV in left panel 

%-- Fig. 2 can be variation with h, fill in more values 

%-- mention what happens when mchi less than TR 

%-- after describing behavior, make a bullet-point list of %main features (TR controls the smallest mX...) 

We would like to note 
%Using Eqs.(3,4), we find that 
that throughout the allowed parameter space shown in Figs.~\ref{fig:TO_mX1},\ref{fig:TO_mX2}. 
%we have 
the expressions given in Eqs.~(\ref{xsection1},\ref{xsection2}) result in $\langle \sigma_{\rm ann} v \rangle \lesssim 10^{-43}$ cm$^3$ s$^{-1}$. The most extreme case in Fig.~\ref{fig:TO_mX3}, corresponding to $h = 10^{-3}$ and $m_X = 1$ TeV, results in $\langle \sigma_{\rm ann} v \rangle \simeq 10^{-37}$ cm$^3$ s$^{-1}$. This is still small enough to be in the freeze-in regime~\cite{Erickcek:2015jza}.  
% \textbf{I'm getting \(10^{-35}\) for the extreme case of \(h=10^{-3}\) and \(m_X=10^2\) with Eq 3. For \(h=10^{-5}\) I get \(10^{-43}\). I think -35 is still freeze-in. {\underline {R.A. I have modified the statement and added a footnote.}}}
%This 
%This is too small to yield the correct relic abundance via %inverse annihilations even when evolution during the %adiabatic phase of the EMD period is taken into %account~\cite{Allahverdi:2019jsc} (details of this %calculation are discussed in Appendix \ref{App:B}). 
It also results in an annihilation rate for $\chi$ that satisfies $\Gamma_{\rm ann} \ll H$ at temperatures $T \lesssim m_\chi$, which ensures that the comoving number density of $\chi$ remains frozen upon production from $X$ decay. In Appendix \ref{App:B}, we discuss DM production from inverse annihilations and find that in large regions of the parameter space it can be completely neglected. In cases with $m_\chi < T_{\rm R}$ and $m_X \gg T_{\rm tr}$, this contribution becomes significant and may even give rise to the correct abundance while $X$ decays lead to DM underproduction.    
%%%%%%%%%%%%%%%%%%%%%%%%%%%%%%%%%%%%%%%%%%%%%%%%%%%%%%%%%%%
\subsection{Scenarios with hidden sector DM}

In this section, we consider the case where \(X\) and \(\chi\) are part of a hidden sector with its own gauge symmetry and temperature that can be distinct from those of the visible sector. The Boltzmann equations in this case are slightly modified to include the radiation energy density of the hidden sector: 
\begin{eqnarray} \label{HS}
\frac{d\rho_{\rm r}^{\rm VS}}{dt} + 4H\rho_{\rm r}^{\rm VS} & = & {\rm Br}_{\phi \rightarrow {\rm VS}}\Gamma_\phi \rho_\phi \, , \nonumber \\
\frac{d\rho_{\rm r}^{\rm HS}}{dt} + 4H\rho_{\rm r}^{\rm HS} & = & \left(1 - {\rm Br}_{\phi \rightarrow {\rm VS}}\right)\Gamma_\phi \rho_\phi \, ,
\end{eqnarray}
where superscripts denote the relevant sector and ${\rm Br}_{\phi \rightarrow {\rm VS}}$ is the branching fraction for \(\phi\) decays to the visible sector. 
In order to guarantee a RD Universe that is compatible with observations, we need to have $\rho^{\rm VS}_{\rm r} \gg \rho^{\rm HS}_{\rm r}$ at the end of EMD. This implies that \(\phi\) must decay predominantly to the visible sector, and hence ${\rm Br}_{\phi \rightarrow {\rm VS}} \simeq 1$.
%Therefore, we take \({\rm Br}_{\phi \rightarrow {\rm VS}} %= 1\) in the following.

There are two major differences between this case and that considered in the previous sections:
\vskip 2mm
\noindent
{\bf (1)} $m_X$ does not need to be restricted to being larger than ${\cal O}({\rm TeV})$. This is because $X$ is a hidden-sector particle, and hence it can easily evade the LHC bounds. All that is required for our analysis to be valid in this case is that $m_X \gg m_\chi$.

\vskip 1.5mm
\noindent
{\bf (2)} Given that $\phi$ decay must mainly reheat the visible sector, the amount of radiation injected to the hidden sector does not necessarily dominate over the initial radiation therein. As a result, the nonadiabatic phase of EMD may be shortened or even non-existent as far as the hidden sector is concerned. 
%This happens when the initial value of $\rho^{\rm %HS}_{\rm r}$ is sufficiently large so that it always %dominates over that fed to the hidden sector by $\phi$ %decay. 
%
\vskip 2mm
In order to solve the Boltzmann equations in Eq.~(\ref{HS}), we need to set the initial values of $\rho^{\rm HS}_{\rm r}$ and $\rho^{\rm VS}_{\rm r}$ as well as the value of ${\rm Br}_{\phi \rightarrow {\rm VS}}$. Here, we consider a case where radiation is almost entirely in the hidden sector prior to EMD. This, for example, is the case in the model of high scale SUSY discussed in~\cite{Allahverdi:2020uax,Allahverdi:2023nov}. We also use the current observational limits from PLANCK 2018 on dark radiation in the form of $\Delta N_{\rm eff}$ using the full TT,TE,EE+lowE+lensing+BAO data~\cite{Planck:2018vyg} to set a lower bound on ${\rm Br}_{\phi \rightarrow {\rm VS}}$. These conditions yield a maximally constraining case for the configuration of the HS in that it is initially dominant and reheated to the maximum amount allowed by observations. An initially subdominant HS, as well as a shorter or even absent nonadiabatic phase, results in weaker constrains in the \(T_{\rm O}\)-\(m_X\) plane. 

\begin{figure}[t!]
\centering
\includegraphics[width = 0.90\textwidth]{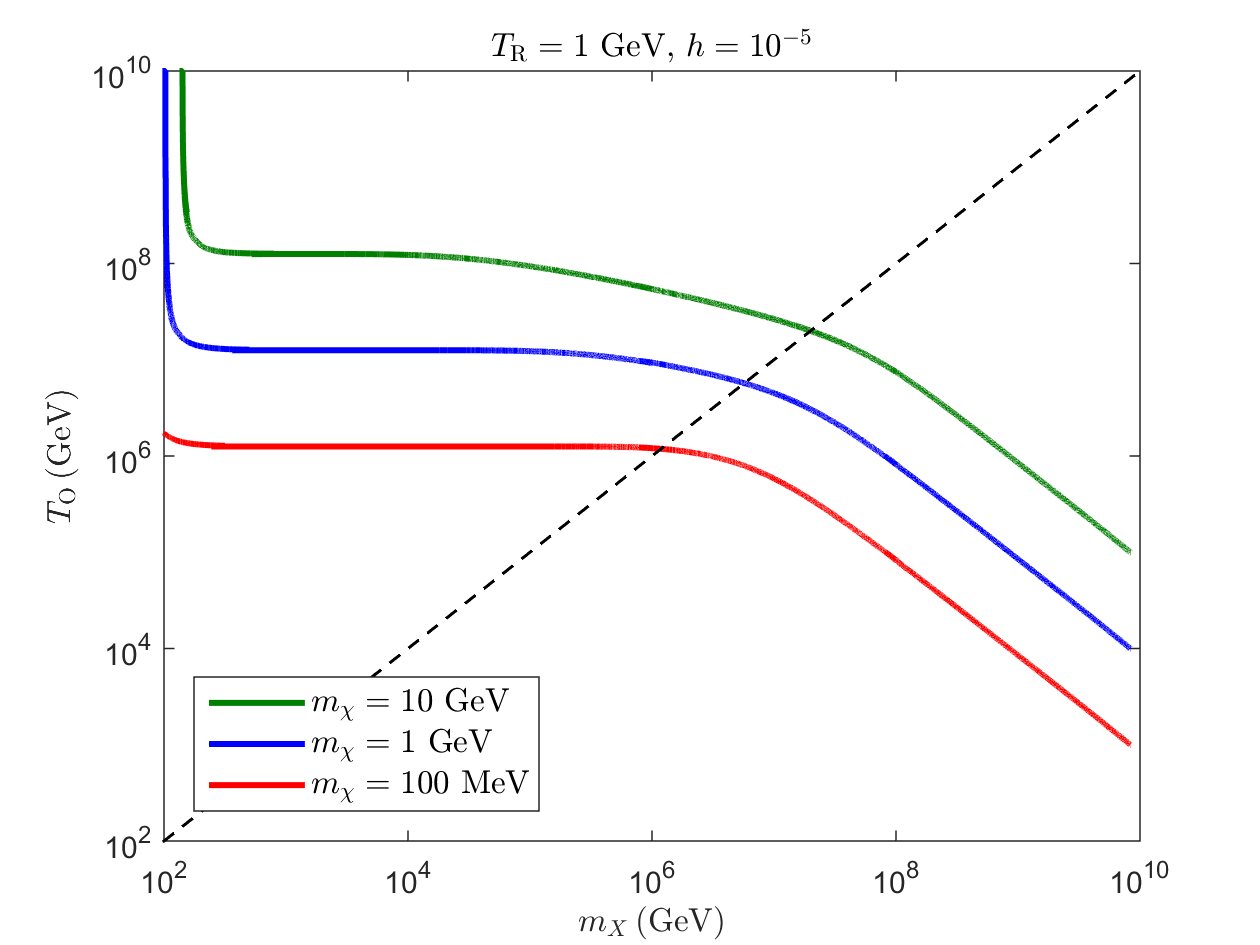}
\caption{The \(T_{\rm O}\)-\(m_X\) plane for varied DM mass \(m_\chi\), with other parameters held fixed, for the case that \(X\) and \(\chi\) are in a hidden sector. The hidden sector is taken to be dominant in the RD period before EMD, and is partially reheated during EMD to the amount allowed by PLANCK 2018 constraints on dark radiation from \(\Delta N_{\rm eff}\). 
% Dashed lines correspond to hidden sector case with hidden radiation initially dominant and \(\phi\) only reheats the visible sector. They begin to curve downward when \(m_X\) approaches \(m_\chi\). The bottom green dashed line is a case where the initial HS temperature is 10 times smaller than that of the VS.
}
\label{fig:10HS}
\end{figure}

In Fig.~\ref{fig:10HS} we show contours for different values of $m_\chi$ that yield the correct DM relic abundance for the benchmark with $T_{\rm R} = 1$ GeV and $h = 10^{-5}$. We note that the vertical segments of the contours have shifted to the left, compared with Fig.~\ref{fig:TO_mX1}, or completely disappeared. This is a consequence of having ${\rm Br}_{\phi \rightarrow {\rm HS}} \equiv 1 - {\rm Br}_{\phi \rightarrow {\rm VS}} \ll 1$. Combined with the removal of the $m_X \gtrsim {\cal O}({\rm TeV})$ constraint, this results in a significant opening up of the allowed parameter space for small values of $m_X$ in comparison with Fig.~\ref{fig:TO_mX1}.      
%\vskip 2mm
%\noindent
%{\bf make HS lines solid, VS dashed, maybe two panels for %different branching.} 

%This, combined with ${\rm Br}_{\phi \rightarrow {\rm VS}} %\simeq 1$, implies an initially subdominant visible %sector that will be populated at the end of EMD
%\footnote{This, for example, is the case in the model of %high scale SUSY discussed in~\cite{Allahverdi:2020uax}.}. 
%Note that the hidden sector strictly remains in an %adiabatic phase throughout the EMD epoch in this case.     

%
%
%-- mention that we have two new parameters in this case, %Br and initial ratio 

%%%%%%%%%%%%%%%%%%%%%%%%%%%%%%%%%%%%%%%%%
\section{Conclusion}\label{sec5}

We presented a detailed study of mediator production and decay and their contribution to the DM relic abundance in scenarios with an epoch of EMD. Special attention was paid to mediators that are charged under the SM gauge group and decay during early stages of EMD (or before its onset).  

We showed that decay of on-shell mediators can totally dominate over the standard freeze-in contribution from inverse annihilations that happen at much lower temperatures. 
%When the lower collider limits of ${\cal O}({\rm TeV})$ %on the mediator mass are taken into account, 
The requirement of not overproducing DM then leads to stringent constraints on the parameter space, demonstrated in Figs.~\ref{fig:TO_mX1}-\ref{fig:TO_mX3}, for DM masses above a few TeV. Additional information on the scale of inflation $H_{\rm inf}$, or an upper bound on it, will further restrict the allowed parameter space in favor of small DM masses. 

%in order to avoid DM overproduction. 
%The allowed regions, shown in Fig.~1-3, typically favor %sub-Tev values of DM mass unless mediators have a mass 
%(well) above $10^{10}$ GeV or for prolonged epochs of EMD %possibly ending just before the onset of BBN. 
%unless mediators are heavier than $10^8$ GeV or so.    
The situation will be more relaxed if the DM candidate and mediators belong to a hidden sector with its own gauge symmetries. The absence of collider bounds on the mediator mass combined with a subdominant reheating of the hidden sector (as required by observational bounds) opens up new regions of the parameter space in this case as seen in Fig.~\ref{fig:10HS}. 
%The EMD epoch may last (much) shorter and it could not %even have a nonadiabatic phase in this case.     

%\textbf{Loc}:
%It is interesting to note that the energy scale of inflation %has to be less than about $10^9$ GeV in order to not allow %sub-Planckian modes to exit the horizon, according to the %recently proposed Trans-Planckian censorship conjecture of %the Swampland program \cite{Bedroya:2019tba}. This %consideration can further constrain our parameter space from %the top (right) in the case when $T_{\rm O}>m_X$ ($T_{\rm O}
%<m_X$).

While mentioning some specific examples of cosmological histories with EMD and DM candidates, we remained largely agnostic about the particle physics origin of DM and the underlying model governing the cosmological history in this work. Our results can therefore be used by both phenomenologists and early Universe model builders to constrain DM models for a given thermal history and vice versa. They can also be used to pinpoint the parameters of a top-down high energy physics model that describes evolution of the early Universe from inflation to BBN and explains DM. 

\section*{Acknowledgments}

The work of R.A. and N.P.D.L. is supported in part by NSF Grant No. PHY-2210367. N.P.D.L. also acknowledges financial support from the UNM Department of Physics and Astronomy through the Origins of the Universe Award. J.K.O. is supported by the project AstroCeNT: Particle Astrophysics Science and Technology Centre, carried out within the International Research Agendas programme of the Foundation for Polish Science financed by the European Union under the European Regional Development Fund. This work was performed in part at Aspen Center for Physics, which is supported by National Science Foundation grant PHY-2210452.

\appendix

%\newpage
\section{DM abundance from \(X\) decay}\label{App:A}

To estimate the final abundance of \(\chi\) particles produced by decay of \(X\) during EMD, we consider the redshifted number density of \(\chi\) from the time that its comoving value becomes frozen, marked by \(T = T_{\rm f}\), to the end of EMD when \(T = T_{\rm R}\), after which point entropy injection ceases: 
\begin{equation}
    n_{\chi}(T_{\rm R}) = n_{\chi}(T_{\rm f}) \left(\frac{a_{\rm f}}{a_{\rm R}}\right)^3 = n_{\chi}(T_{\rm f}) \left(\frac{H_{\rm R}}{H_{\rm f}}\right)^2 \,,
\end{equation}
with 
\begin{equation}
    n_{\chi}(T_{\rm f}) = F(\gamma_\chi) n_{\chi}^{\rm eq}(T_{\rm f}) = F(\gamma_\chi) \frac{\zeta(3)}{\pi^2} g_\chi T_{\rm f}^3 \,,
\end{equation} 
assuming \(T \gg m_\chi\). 
% and \(T_{\rm f} \approx m_X\) (can also do \(m_X/5\) or \(m_X/10\) to be more precise). 
Taking \(T_{\rm R}\) to be when \(\rho_\phi = \rho_{\rm r}\) at the end of EMD, to better match numerical results, we have 
\begin{equation}
    H_{\rm R} = \left(\frac{\pi^2 g_{*\rm R}}{45}\right)^{1/2}\frac{T_{\rm R}^2}{M_{\rm P}} \,.
\end{equation}

\noindent
{\bf (1)} \(T_{\rm tr} < T_{\rm f} < T_{\rm O}\) - In this case the comoving \(\chi\) number density freezes during the adiabatic phase and the Hubble rate at that time is therefore given by: 
\begin{equation}
    H_{\rm f} = \left(\frac{\pi^2}{45}g_{*\rm f}\right)^{1/2} \frac{T_{\rm f}^{3/2} T_{\rm O}^{1/2}}{M_{\rm P}} \,,
\end{equation}
where we have used a factor of 2 for \(\rho_\phi(T_{\rm O}) = \rho_{\rm r}(T_{\rm O})\), and \(g_*T^3 \propto a^{-3}\). 
After normalizing by the entropy density at \(T_{\rm R}\), this gives: 
% \begin{equation}
%     n_{\chi{\rm R}} = F(\gamma_\chi) \frac{\zeta(3)}{\pi^2} g_\chi T_{\rm f}^3 \left(\frac{\left(\frac{\pi^2 g_{*\rm R}}{45}\right)^{1/2}\frac{T_{\rm R}^2}{M_{\rm p}}}{\left(\frac{\pi^2}{45}g_{*f}\right)^{1/2} \frac{T_{\rm f}^{3/2} T_{\rm O}^{1/2}}{M_{\rm p}}}\right)^2 
%     = F(\gamma_\chi) \frac{\zeta(3)}{\pi^2} g_\chi \left(\frac{g_{*\rm R}}{g_{*f}}\right) \left(\frac{T_{\rm R}^4}{T_{\rm O}}\right)
% \end{equation}
% and 
\begin{equation} \label{decA}
    \left(\frac{n_\chi}{s}\right)_{\rm dec}^{\rm A} \approx 0.28 F(\gamma_\chi) \frac{g_\chi}{g_{*\rm f}} \left(\frac{T_{\rm R}}{T_{\rm O}}\right) \,.
\end{equation}

\noindent
{\bf (2)} \(T_{\rm R} < T_{\rm f} < T_{\rm tr}\) - The number density freezes during the nonadiabatic phase and the corresponding Hubble rate is given by: 
\begin{equation}
    H_{\rm f} = \frac{1}{\sqrt{6}} \left(\frac{5\pi g_{*\rm f}}{6\sqrt{10} g_{*\rm R}^{1/2}}\right) \frac{T_{\rm f}^4}{T_{\rm R}^2 M_{\rm P}} \,,
\end{equation} 
where we have taken \(\Gamma_\phi = \sqrt{3} H_{\rm R}\) to match our numerical results. After normalization, this gives: 
% \begin{equation}
%     n_{\chi{\rm R}} = F(\gamma_\chi) \frac{\zeta(3)}{\pi^2} g_\chi T_{\rm f}^3 \left(\frac{\left(\frac{\pi^2 g_{*\rm R}}{45}\right)^{1/2}\frac{T_{\rm R}^2}{M_{\rm p}}}{\frac{1}{A\sqrt{2}} \left(\frac{5\pi g_{*f}}{6\sqrt{10} g_{*R}^{1/2}}\right) \frac{T_{\rm f}^4}{T_{\rm R}^2 M_{\rm p}}}\right)^2 
%     = F(\gamma_\chi) \frac{16\zeta(3)g_\chi A^2}{25\pi^2} \left(\frac{g_{*\rm R}}{g_{*f}}\right)^2 \left(\frac{T_{\rm R}^8}{T_{\rm f}^5}\right)
% \end{equation}
% and 
\begin{equation}\label{decNA}
    \left(\frac{n_\chi}{s}\right)_{\rm dec}^{\rm NA} \approx 0.53 F(\gamma_\chi) \frac{g_\chi g_{*\rm R}}{g_{*\rm f}^2} \left(\frac{T_{\rm R}}{T_{\rm f}}\right)^5 \,.
\end{equation}

\section{Production from inverse annihilations}\label{App:B}

Freeze-in production of DM from inverse annihilations can happen in three different regimes:
\vskip 2mm
\noindent
{\bf (1)} $m_\chi > T_{\rm tr}$. In this regime, inverse annihilations during the adiabatic phase of the EMD epoch (or whatever phase precedes it) will be the dominant source. In fact, as shown in~\cite{Allahverdi:2019jsc}, the main contribution comes from the highest temperature at which the expressions in Eqs.~(\ref{xsection1},\ref{xsection2}) remain valid, namely $T \sim m_X$. The relic abundance from inverse annihilations with \(m_X < T_{\rm O}\) then follows~\cite{Allahverdi:2022zqr,Allahverdi:2019jsc}
\begin{eqnarray} \label{A2}
%&& 
%\left(\Omega_\chi h^2\right)^{\rm A} \simeq \frac{1.3 %\times 10^8} {g_{*m_X}^{3/2}} \left({g_\chi \over %2}\right)^2\left({T_{\rm R} \over 1 ~ {\rm GeV}}\right) %\left(\frac{m_\chi}{100\,{\rm GeV}}\right) \left({m_X \over %T_{\rm O}}\right)^{3/2} \left(\frac{\left<\sigma_{\rm ann} %v\right> (T \sim m_X)}{10^{-36}\,{\rm cm}^3\,{\rm %s}^{-1}}\right)  
%\, , \nonumber \\ 
%&& 
\left({n_\chi \over s} \right)_{\rm ann}^{\rm A} \sim 10^{-1} \frac{g_\chi}{g_{*m_X}^{3/2}} \langle \sigma_{\rm ann} v \rangle (E \sim m_X) M_{\rm P} T_{\rm R} \left({m_X \over T_{\rm O}}\right)^{3/2} 
\, .
\end{eqnarray}
If instead \(T_{\rm O} < m_X\), one can obtain a similar expression based on the type of dominant energy density before EMD. 
% , the abundance is given by 
% \begin{eqnarray} \label{A2}
% && 
% \left(\Omega_\chi h^2\right)^{\rm A} \simeq \frac{2.6 \times 10^9} {g_{*m_X}^{3/2}} \left({g_\chi \over 2}\right)^2\left({T_{\rm R} \over 1 ~ {\rm GeV}}\right) \left(\frac{m_\chi}{100\,{\rm GeV}}\right) \left({m_X \over T_{\rm O}}\right) \left(\frac{\left<\sigma_{\rm ann} v\right> (T \sim m_X)}{10^{-36}\,{\rm cm}^3\,{\rm s}^{-1}}\right) 
% %~, ~ ~ ~ ~     \langle \sigma_{\rm ann} v \rangle < %\frac{\pi^3 g_{*\rm O}^{1/2}}{\sqrt{90} \zeta(3) M_{\rm p} %T_{\rm O}} 
% %\lesssim \left<\sigma v\right>^{\rm e-eq}, 
% \, , \nonumber \\ 
% && \left({n_\chi \over s} \right)_{\rm ann}^{\rm A} \sim 10^{-4} \langle \sigma_{\rm ann} v \rangle (E \sim m_X) M_{\rm P} T_{\rm R} \left({m_X \over T_{\rm O}}\right) 
% %\left(\Omega_\chi h^2\right)^{\rm A} \simeq \frac{0.076}
% %{g_{*\rm O}} \left(\frac{10^9\,T_{\rm R}}{T_{\rm O}}\right) %\left(\frac{m_\chi}{1\,{\rm GeV}}\right) ~, ~ ~ ~ ~ ~ ~ ~ ~ %~ ~ ~ ~ ~ ~ ~  \langle \sigma_{\rm ann} v \rangle \gtrsim %\frac{\pi^3 g_{*\rm O}^{1/2}}{\sqrt{90} \zeta(3) M_{\rm p} %T_{\rm O}} 
% \, .
% \end{eqnarray}
%
\vskip 1.5mm
\noindent
{\bf (2)} $T_{\rm R} < m_\chi \lesssim T_{\rm tr}$. In this case, the main contribution from inverse annihilations arises in the nonadiabatic phase of EMD (at $T \sim m_\chi/4$) giving rise to the following (for example, see~\cite{Erickcek:2015jza}): 
%paper, freeze-in abundance during nonadiabatic EMD (\%%(T_{\rm R} \lesssim m_\chi/4 \lesssim T_{\rm tr}\)) due to \
%(2 \to 2\) scattering is given by: 
%
\begin{eqnarray} \label{NA2}
%&& \left(\Omega_\chi h^2\right)^{\rm NA} \simeq 8.4 \times %10^7 \frac{g_{*\rm R}^{3/2}}{g_{*\rm m_\chi /4}^3} %\left(\frac{g_\chi}{2}\right)^2 \left(\frac{T_{\rm R}}
%{m_\chi}\right)^5 \left(\frac{T_{\rm R}}{1\,{\rm %GeV}}\right)^2 \left(\frac{\left<\sigma_{\rm ann} v\right> 
%(T \sim m_\chi/4)}{10^{-36}\,{\rm cm}^3\,{\rm %s}^{-1}}\right) \, , \nonumber \\
%&& 
\left({n_\chi \over s} \right)_{\rm ann}^{\rm NA} \sim \frac{g_{*\rm R}^{3/2} g_\chi}{g_{*\rm m_\chi /4}^3} \langle \sigma_{\rm ann} v \rangle (E \sim m_\chi/4) M_{\rm P} T_{\rm R} \left({T_{\rm R} \over m_\chi}\right)^6 \, . 
\end{eqnarray}
\vskip 1.5mm
\noindent
{\bf (3)} $m_\chi \lesssim T_{\rm R}$. In this case, production from inverse annihilations is most efficient in the RD phase after EMD. The abundance is set at \(T \sim T_{\rm R}\). %Here the abundance 
and can be estimated as:
\begin{eqnarray} \label{R2}
%&& \left(\Omega_\chi h^2\right)^{\rm RD} \simeq \frac{2.3 %\times 10^{9}}{g_{*\rm R}^{3/2}} \left(\frac{g_\chi}
%{2}\right)^2 \left(\frac{T_{\rm R}}{1\,{\rm GeV}}\right) %\left(\frac{m_\chi}{100\,{\rm GeV}}\right) %\left(\frac{\left<\sigma_{\rm ann} v\right> (T \sim T_{\rm %R})}{10^{-36}\,{\rm cm}^3\,{\rm s}^{-1}}\right) \, , %\nonumber \\
%&& 
\left({n_\chi \over s} \right)_{\rm ann}^{\rm RD} \sim 10^{-1} \frac{g_\chi}{g_{*\rm R}^{3/2}} \langle \sigma_{\rm ann} v \rangle (E \sim T_{\rm R}) M_{\rm P} T_{\rm R} \, .
\end{eqnarray}
%
%\begin{equation}
%    = \frac{2.7\times 10^{26}}{g_{*\rm R}^{3/2}} %\left(\frac{g_\chi}{2}\right)^2 T_{\rm R} m_\chi \frac{C %h^4}{16 \pi m_X^2} \left[1,\frac{m_\chi}{m_X}\right]^2 
%\end{equation}
%
%\vskip 2mm
%The values of $T_{\rm R}$, $T_{\rm O}$, and \(m_\chi\) %determine which of the expressions in Eqs~
%(\ref{NA2},\ref{A2},\ref{R2}) applies. 
Note that the value of $\langle \sigma_{\rm ann} v \rangle$ at the relevant temperature in each case should be used, which is important when it is not a constant, see Eq.~(\ref{xsection2}).
%to the contribution of inverse annihilations to the DM relic %abundance. 
%
\vskip 2mm
The values of $m_X$, $m_\chi$, $T_{\rm R}$, and $T_{\rm O}$ determine the relevant regimes for DM production from inverse annihilations and $X$ decay (discussed in Appendix \ref{App:A}).  
To compare the contribution of inverse annihilations with that from $X$ decay, we express \(\left<\sigma_{\rm ann} v\right>\) in terms of the particle-physics parameters using Eqs.~(\ref{xsection1},\ref{xsection2}) as an example. 
% 
%It is easy to see that $X$ decay totally dominates when %$T_{\rm tr} < m_\chi < m_X$, in which case we should compare %the expressions in Eqs.~(\ref{decA}) and (\ref{A2}). This is %expected as both contributions mainly come from the epoch %when $T \sim m_X$ and annihilations are suppressed by a %higher power of the coupling $h$. 
Inverse annihilations become most pronounced in the regime where $m_\chi < T_{\rm R}$ and $m_X > T_{\rm tr}$. In this case, the number density of DM particles produced from $X$ decay has undergone the highest redshift by the time inverse annihilations become efficient.      
For the parameter combinations shown in Figs.~\ref{fig:TO_mX1}-\ref{fig:TO_mX3}, we find that inverse annihilations contribute more than the observed DM abundance only in the regions where decay production also overproduces DM, with one exception. For \(h = 10^{-3}\) in Fig.~\ref{fig:TO_mX3}, inverse annihilations in the nonadiabatic phase lead to \(\Omega_\chi h^2 \sim 0.1\) for \(m_X \approx 1\)~TeV and large values of \(T_{\rm O}\), essentially coinciding with the contour from decay production. For larger values of \(m_X\) the abundance from inverse annihilation decreases.

\bibliographystyle{JHEP}
\bibliography{references}

\end{document}